\documentclass[graybox,natbib,nosecnum]{svmult}
\bibpunct{(}{)}{;}{a}{}{,} % suppress commas between author-names and year

\pdfoutput=1   %forces use of pdflatex. Disable if you prefer to use .eps or .ps figures.
\usepackage{mathptmx}       % selects Times Roman as basic font
\usepackage{helvet}         % selects Helvetica as sans-serif font
\usepackage{courier}        % selects Courier as typewriter font
\usepackage{type1cm}        % activate if the above 3 fonts are
\usepackage{makeidx}         % allows index generation
\usepackage{graphicx}        % standard LaTeX graphics tool
\usepackage{multicol}        % used for the two-column index
\usepackage[bottom]{footmisc}% places footnotes at page bottom
\usepackage[normalem]{ulem}	% for strike-through of text with \sout{}  
\usepackage{hyperref}  %for hyperlinks

\usepackage{soul}   % for high-lighting of text
\makeindex             % used for the subject index
\begin{document}

\title*{Galactic Effects on Habitability}
\author{Nathan A. Kaib}
\institute{Nathan A. Kaib \at HL Dodge Department of Physics and Astronomy, University of Oklahoma, Norman, OK, USA, \email{nathan.kaib@ou.edu}}
\maketitle

\abstract{The local galactic environment has been suspected to influence planetary habitability in numerous ways. Very metal-poor regions of the Galaxy, or those largely devoid of atoms more massive than H and He, are thought to be unable to form habitable planets. Moreover, if such planets do form, the newly formed system is subjected to close stellar passages while it still resides in its stellar birth cluster. After star clusters disperse, various potential hazards still remain. For instance, the central galactic regions may present risks to planetary habitability via nearby supernovae, gamma ray bursts (GRBs), and frequent comet showers. In addition, planets residing within very wide binary star systems are affected by the Galaxy, as local gravitational perturbations from the Galaxy can increase the binary's eccentricity until it destabilizes the planets it hosts. Here we review the most recent work on the main galactic influences over planetary habitability. Although there must be some metallicity limit below which rocky planets cannot form, recent exoplanet surveys show that they form around stars with a very large range of metallicities. Once formed, the probability of star cluster environments destabilizing planetary systems only becomes high for rare, extremely long-lived clusters. Regarding the threats to habitability from supernovae, GRBs, and comet showers, many recent studies of these processes suggest that their hazards are more limited than originally thought. Finally, denser regions of the Galaxy will enhance the threat that very wide binary companions pose to planetary habitability, but the probability that a very wide binary star will disrupt habitability will always be substantially below 100\% for any galactic environment. While some regions of the Milky Way must be more hospitable to habitable planets than others, it is very difficult to state that habitable planets are confined to any well-defined region of the Galaxy or that any other particular region of the Galaxy is completely devoid of habitable planets.}

\section{Introduction }

Because the typical distances between stars are so great, it is tempting to think of planetary systems as self-contained, isolated systems separate from the rest of the Galaxy. For many considerations, this picture is perfectly valid. However, there is a level of connection between a planetary system and its local galactic environment, and this can sometimes have dramatic consequences for the formation and evolution of planetary systems. Because the processes of planetary formation and evolution are strongly tied to habitability, there is also a link between the Galaxy and planetary habitability. Specifically, the metallicity of the local interstellar medium (ISM) should influence the efficiency of terrestrial planet formation around stars by setting the mass of solids in their protoplanetary disks \citep[e.g. ][]{liss95}. Moreover, after planet formation is complete, planetary systems can be destabilized by close encounters with other stars while they still inhabit their crowded birth clusters \citep{mott77, gaid95, adamslaugh01}. After leaving their birth clusters, they can exposed to (and possibly sterilized by) nearby energetic events such as supernovae \citep{ell95} and gamma ray bursts, or GRBs \citep{thorsett95}. In addition, there are the gravitational effects of the Galaxy to consider. The most distant portion of our own solar system, the Oort cloud, is continually perturbed by the gravity of passing field stars as well as the tide of the Milky Way's disk \citep{oort50, heitre86, hei87}. By driving orbits of Oort Cloud objects to very high eccentricities, both of these perturbations place long-period comets on orbits that pass near (and potentially impact) Earth \citep{hills81, mat95, kaibquinn09}. In a very similar manner to Oort cloud objects, these same galactic perturbations can drive the orbits of very wide binary stars through phases of very high eccentricity \citep{jiangtre10, kaib14}, and the typical very wide binary will pass through one or more of these brief high eccentricity phases \citep{kaib13}. This process can make very wide stellar companions quite hazardous to the stability of planetary systems, since the planets can spend Gyrs with minimal gravitational interactions with the star before undergoing numerous close encounters with the stellar companion when it attains a high orbital eccentricity \citep{kaib13}. Thus, a range of different processes must be considered when thinking about the Galaxy's influence over planetary habitability. With this in mind, we will review each one of these processes in detail in the sections below and assess how strongly it influences planetary habitability.

\section{Galactic Metallicity}

The first exoplanets ever discovered around main sequence stars were Jovian mass bodies \citep[e.g., ][]{may95}. After the first few discoveries, it was quickly realized that metal-rich stars (stars enriched with atoms more massive than H and He) are much more likely to host these types of planets than their metal-poor counterparts \citep{gon97}. While this effect was initially thought to be due to metal-pollution from accreted planetary material, it has since been definitively shown instead that a high primordial metallicity of the proto-star enhances the formation of giant planets \citep[e.g., ][]{gon97, gon98, sant03}. \citet{fischval05} found that stars with a metallicity 0.3 dex or more above solar metallicity were $\sim$10 times more likely to host a giant planet than stars with metallicities 0.5 dex or more below the Sun's. On the surface, this result suggests that the overall presence of planets (and therefore habitable planets) is strongly dependent on the host star's metallicity, which is of course determined by the metallicity of the Galaxy's interstellar medium (ISM) \citep[e.g.][]{mattfran89}. Since the Milky Way's ISM metallicity generally falls off with galactocentric distance, this would in turn imply that planets should be relatively common near the Galactic center, while the outer regions of the Milky Way's disk should be largely devoid of planets \citep{gon01, line04}.  

However, it is less clear there is a strong correlation between stellar metallicity and the prevalence of less massive planets. Using the catalog of exoplanet candidates discovered by the Kepler mission \citep{borucki10}, \citet{buch12} found that planets with radii below 4 R$_{\oplus}$ are found around a wide range of stellar metallicities unexpected by extrapolation of the giant planet metallicity dependence down to lower masses. Followup work continued to find a lack of evidence for a metallicity dependence on low-mass planet formation \citep{buch15}. However, some of these works may suffer from systematic errors \citep{zhu16}. Moreover, lower mass planets generally seem to be much more common than high-mass planets, and this combined with the difficulty of their detection may hide any correlation between their formation and stellar metallicity \citep{zhu16}. Nevertheless, at a minimum it remains unclear how strongly dependent the prevalence of terrestrial mass planets is on stellar metallicity and, hence, the Galaxy's metallicity gradient. 

Another important consideration is the key role that plate tectonics has played in maintaining Earth's habitability. The concept of a greenhouse effect regulated by geological activity is thought to be a requirement for long-term climate stability \citep{kast93}, and a large portion of the internal heat that drives Earth's plate tectonics is thought to be generated from the decay of $^{235,238}$U, $^{232}$Th, and $^{40}$K \citep[e.g., ][]{fowler90, kor11}. Historically, Th and U have traditionally been thought to originate during type-II supernovae \citep{burb57, woo86}. Thus, maintaining a steady supply of these unstable isotopes in the ISM seemed to require regular type-II supernovae, in turn requiring recent star formation. Moreover, although K can be synthesized through several processes \citep[e.g., ][]{busso99, her05}, its dispersal into the ISM likely mainly takes place through supernovae as well. Under this framework, as star formation falls off in the Milky Way over time, the injection rate of these elements into the ISM also falls off, while iron's injection rate does not change as much \citep{mcwill97}. Consequently, this predicts that on average, terrestrial planets on average should have a smaller mass fraction of radiogenic isotopes if they are formed at late epochs when the galactic star formation rate is lower compared to planets formed during earlier epochs with a higher star formation rate \citep{gon01}. This suggests that the ability to form planets with plate tectonics may be dependent on the Galaxy's star formation history. 

However, neutron star mergers, which require the existence of massive stars but are less temporally tied to their life cycles, are also a proposed pathway for U and Th production \citep{eich89, frei99}. The fist observation of such an event through the detections of both the optical and gravitational wave signatures has confirmed that neutron star mergers are a major, and likely the dominant, contributor to r-process elements, including U and Th \citep[e.g.][]{abb17a, abb17b, abb17c, cow17}. Thus, an ISM enriched in these unstable isotopes may not require extremely recent star formation. Moreover, the presumed dependence of plate tectonics on internal heat fueled by radioactive decay neglects the contribution of the primordial heat generated during the formation of the Earth. While there is a large amount of uncertainty in the relative contributions of Earth's sources of internal heat, it appears that the heat flux from radioactive decay is roughly equal to the flux from the Earth's primordial heat \citep{kam11, kor08}. Thus, even without a large supply of radiogenic heat, the heat flow at the modern Earth's surface would still be within a factor of $\sim$2 of what is observed today. 

\section{Star Cluster Phase}

The vast majority of stars (and therefore planets) form within star clusters \citep[e.g.][]{lala03}. Before these clusters disperse, their stars and planets inhabit environments much denser than the typical galactic field star. The enhanced stellar densities and the lower relative velocities of stars within clusters lead to star-star encounters that are much closer and more powerful than those experienced in the galactic field \citep{mott77, fern97}. This then raises the possibility that the stellar encounters experienced by a planetary system during the cluster phase may be powerful enough to destabilize the planets' orbits \citep{adamslaugh01}. Although the small orbital semimajor axis of a planet similar to the Earth would require an incredibly close stellar encounter to be directly perturbed \citep{laughadams00}, a much more modest stellar encounter could excite the orbit of an outer planet such as Neptune, and this excitation can cascade in toward habitable planets via planet-planet interactions \citep{adamslaugh01, zaktre04, malm11}. 

Assuming a reasonable distribution of stellar masses \citep{adamsfatuzzo96} and a typical cluster stellar encounter velocity of $\sim$1 km/s, \citet{adamslaugh01} find that doubling the orbital eccentricity of a Neptune-like planet would require a stellar passage within $\sim$225 AU.  To completely destabilize a system of planets like our own giant planets requires an even closer encounter inside $\sim$100 AU \citep{malm11}. If we assume the planets' parent star inhabited a star cluster with a particularly high density of $10^3$ stars/pc$^3$, then the system would have to remain in the cluster for $\sim$250 Myrs before an encounter within $\sim$225 AU would be expected \citep{adams10}. Of course, an encounter within 100 AU would require an even longer residence time. Meanwhile, the large majority of stars are actually formed in embedded clusters, which disperse on timescales less than $\sim$10 Myrs \citep{lala03}. Only open clusters, which remain gravitationally bound after their molecular gas disperses, survive for timescales longer than tens of Myrs, and it appears that no more than $\sim$10\% of stars are born into open clusters \citep[e.g.][]{adamsmyers01}. Thus, it seems that if our solar system's architecture is typical of systems that host a habitable world, then the large majority of habitable planets should be safe from dynamical disruption during their cluster phase.

\section{Supernovae}

Once a star leaves its cluster, it transitions to a much less crowded galactic environment. However, even in such rarified environments, it remains possible for the galactic environment to influence habitable planets. One potential way is through high-energy events such as supernovae. Works on the potentially damaging effects of a nearby supernova on Earth's biosphere have appeared in the literature for many decades \citep[e.g., ][]{tertuck68, rud74, brak81, crutbru96}. One particularly biologically significant effect on the modern Earth would be that high-energy photons released during the supernova event would deplete the Earth's ozone layer \citep{rud74}. As a result the surface of the Earth would subsequently be exposed to a greater fraction of the Sun's UV radiation, resulting in a less hospitable environment for life. In the original calculations, \citet{rud74} found that a supernova within 17 pc of Earth would result in an 80\% reduction in Earth's ozone layer that would persist for $\sim$2 years. Moreover, a $\sim$50\% reduction could persist for centuries afterward due to the destructive effects of cosmic rays also generated in the explosion. However, since that initial work, atmospheric modeling has improved dramatically as well as our understanding of supernovae and their radiation spectra. More recent estimates of the effect on the Earth's ozone layer find that the solar UV flux on Earth's surface would only double if a supernova occurred within 8 pc of the Sun \citep{geh03}. While there is evidence that supernovae have occurred within 100 pc of the Earth in the last 10 Myrs \citep{knie99, erlwolf10, breit16}, supernovae within 8 pc are thought to occur of order once per Gyr. Thus, for a galactic environment like the Sun's significant alterations of planetary habitability by supernovae should be relatively rare.

\section{Gamma Ray Bursts}

Even more energetic than supernovae are gamma ray bursts (GRBs), beamed, transient bursts of intense $\gamma$-rays. There are numerous types of GRBs of differing duration, and long GRBs ($t>\sim$2 s) are believed to pose the greatest hazard to life based on their rate and high luminosities \citep{piranjimenez14}. Analogous to supernovae, if a GRB occurs near enough to Earth, the high temporary flux of $\gamma$-rays can deplete the ozone layer and expose the biosphere to harsh UV radiation from the Sun \citep{thorsett95}. In fact, a typical luminosity GRB within 1--2 kpc of Earth may cause an ozone depletion of $\sim$35\%, which may be enough to trigger a mass extinction, one perhaps similar to the Ordivician-Silurian event \citep{thomas05a, thomas05b}. Using the observed GRB luminosity function and occurrence rates observed in the local universe, \citet{piranjimenez14} estimate a $\sim$50\% chance that the Earth has been exposed to a lethal GRB in the past 1 Gyr. Similarly, \citet{lizhang15} estimate that $\sim$1 lethal GRB has occurred near Earth in the last 500 Myrs. 

However, studies on the biological hazards of GRBs still suffer from the uncertainty about the nature of GRBs themselves. The most widely employed model for the GRB progenitor is a collapsing, rapidly spinning high-mass star \citep{macfadyenwoosley99}. The collapse of such a star into a black hole is thought to result in two relativistic jets briefly beaming $\gamma$-rays across the Universe. Given this, it is expected that the rate of GRBs should scale with the star formation rate, which is generally observed \citep{jimenezpiran13}. However, the local GRB rate also seems to have a metallicity dependence wherein most nearby GRBs occur in galaxies with metallicities less then 10\% of the Sun's metallicity \citep{jimenezpiran13}. If it is assumed that the progenitor stars of GRBs must be approximately this metal-poor, \citet{gowanlock16} find that very few stars in the Milky Way should have been sterilized by GRBs in the last Gyr, and these stars should be preferentially located in the sparsely populated outskirts of the Galaxy. Thus, the threat of a GRB-induced mass extinction is likely a scaling probability depending on the local galactic star formation rate and metallicity, but the magnitude and nature of the scaling remains uncertain due to our poor understanding of GRB progenitors. 

\section{Comet Bombardment}

Another way the galactic environment can influence planetary habitability is simply through its gravity. The gravity of the Milky Way's disk is manifested as a tidal field (in the vertical direction relative to the disk midplane) over small distance scales \citep{heitre86}. Moreover, when a field star passes near a planetary system, it can impart a velocity impulse on orbiting bodies relative to their parent star \citep{opik32}. For orbits within tens of AU of the Sun, perturbations from the Galaxy will be inconsequential \citep{laughadams00}. However, the solar system's Oort cloud extends tens of thousands of AU from the Sun \citep{oort50}. At these huge orbital distances the tidal and stellar perturbations from the local galactic environment are the main driver of orbital evolution \citep{dun87}. Over time these perturbations isotropize inclinations and drive a diffusion in semimajor axis \citep{weiss96}. Most importantly, perturbations from the Galaxy drive a pseudo-random walk in the perihelia of Oort cloud objects \citep{heitre86}. In fact, this is how long-period comets are driven into extremely eccentric orbits that take them near the Sun and Earth \citep{oort50, wietre99}, occasionally resulting in impacts on Earth \citep{weiss07}. 

Historically, it was thought that during most times, only Oort cloud objects in the cloud's outer periphery could evolve to potentially Earth-impacting orbits \citep{heitre86}. The reasoning for this is that the perihelia of distant Oort cloud objects evolve quickly under the perturbations of the Galaxy, whereas the Galaxy's perturbations are normally too weak to produce rapid changes in the orbits of more tightly bound Oort cloud objects \citep{hills81, heitre86, dun87}. As a result, the perihelia of more tightly bound Oort cloud objects cannot evolve from beyond the gas giants' orbits to inside the Earth's orbit in a single revolution about the Sun. Because of this, these more tightly bound Oort cloud objects will make perihelion passages in the Jupiter-Saturn zone of the solar system before they ever attain a perihelion near Earth. During perihelion passages in the Jupiter-Saturn zone, the gas giants will deliver energy kicks to the Oort cloud objects that will ultimately drive their ejection to interstellar space, preventing them from ever approaching Earth. Thus, it was thought that the modern catalog of LPCs originated exclusively from the areas of the Oort cloud further than 20,000 AU from the Sun.

The only exception to this process was thought to occur during comet showers \citep{hills81, hut87}. These are caused by rare, extremely powerful encounters with field stars. During such an encounter, the perturbation on the Oort cloud is so strong that any object is temporarily able to quickly circumvent the Jupiter-Saturn zone and evolve to an Earth-crossing orbit. Thus, during these rare events, the Earth can be exposed to potential impactors from the inner regions of the Oort cloud and not just the outer periphery. Although the huge orbital periods and isotropic  inclinations of long-period comets make them inefficient Earth impactors \citep{weiss07}, the number of impacting long-period comets could increase by orders of magnitude for a few Myrs during a comet shower, potentially triggering a mass extinction \citep{hut87}. However, one of the main uncertainties of the importance of comet showers is the population of the inner region of the Oort cloud. If there are few objects orbiting in this area of the solar system, then comet showers cannot result in many impacts on Earth. However, if there are many more objects in the inner 20,000 AU of the Oort cloud than the more distant areas, then the effects of comet showers on Earth could be quite severe. 

Regarding the uncertainty in the Oort cloud population size, significant progress was made in \citet{kaibquinn09}, when it was discovered that objects in the inner 20,000 AU of the Oort cloud could in fact evolve to become long-period comets that pass near Earth during non-comet shower times. This occurs because before these slowly evolving orbits receive energy kicks from the planets that are strong enough to eject them, some of them first receive energy kicks that are only strong enough to inflate their semimajor axes above $\sim$20,000 AU. With such large semimajor axes, the perihelia of these objects can slide across the Jupiter-Saturn zone toward Earth during their subsequent revolution about the Sun. An example of this process is shown in Figure 1. Here an object with an initial semimajor axis of $\sim$6,000 AU evolves to an orbital perihelion of $\sim$17 AU. By the time this object is observed from Earth, its semimajor axis has been altered by the planets to $\sim$30,000, disguising its region of origin in the Oort cloud. 

Although the dynamical pathway from the inner region of the Oort cloud is less efficient than that from the outer periphery of the cloud \citep{fouch14}, the modern catalog of long-period comets can still be used to place an upper limit on the number of bodies in the inner region of the Oort cloud. \citet{kaibquinn09} estimated that no more than $\sim$$10^{12}$ comet-sized objects should reside in the entire Oort cloud. Given this upper limit combined with the stellar kinematics of the solar neighborhood, \citet{kaibquinn09} argued that the most powerful comet shower expected to have occurred since the Cambrian Explosion would be unlikely to trigger a mass extinction. Thus, as with supernovae, comet showers are unlikely to have had a significant effect on Earth's habitability. 

\section{Very Wide Binary Stars}

An interesting offshoot from the study of mass extinctions and comet showers was the proposal of the Nemesis hypothesis \citep{whit84, davis84}. This hypothesis argued that the Sun possesses an unseen red dwarf or brown dwarf companion on an extremely wide, but very eccentric orbit. Each time the companion makes a periastron passage, it delivers a severe perturbation to the Oort cloud, triggering a mass extinction-causing comet shower on Earth. 

Modern all-sky surveys have effectively ruled out the existence of such a massive companion to the Sun \citep[e.g., ][]{kirk11}, but $\sim$10\% of other Sun-like stars are actually members of very wide binary star systems, or gravitationally bound stellar pairs separated by 10$^3$ AU or more \citep{dhit10, lep07, long10}. While we have no idea if these systems possess Oort clouds like our own, we do know that many stars, perhaps most, possess planets \citep[e.g., ][]{tuomi14, dresscharb13, petigura13}. Considering the direct interaction between planets and a very wide binary companion raises another potential consequence of a Nemesis-like companion. If its orbit is eccentric enough, consequences even more dramatic than comet showers will result from its pericenter passages. Models of our Sun's giant planets indicates that slow stellar flybys within $\sim$100 AU will deliver such a strong gravitational perturbation to the Sun's outer planets that they will be destabilized \citep{malm11}. This typically results in an episode of planet-planet scattering in which one or more planets are ejected and the orbital eccentricities of the survivors are greatly increased. Very wide binary companions therefore have the potential to alter the architectures of planetary systems via instabilities.

A planetary instability triggered by a very wide binary companion thus requires a companion with a periastron of $\sim$100 AU or lower, or an eccentricity of 0.9 or higher (assuming a planetary architecture like the Sun's). For any dynamically sculpted distribution of stellar orbits, we would expect that only a very tiny fraction of very wide binary stars would possess such extreme eccentricities at any given time. However, from studies of the Sun's Oort cloud, we also know that passing field stars and the tide of the Milky Way perturb weakly bound orbits, and the same processes acting on the Oort cloud will drive the dynamics of very wide binary stars \citep{jiangtre10}. We therefore expect that the periastra (and eccentricities) of very wide binary stars will continually change under perturbations from the Milky Way. Consequently, very wide binaries that have very eccentric orbits can evolve towards circular orbits (and back), and very wide binaries that have circular orbits can evolve to states of high eccentricities (and back as well). 

This raises the prospect that many planetary systems within very wide binary stars may form when the very wide binary has a low or moderate eccentricity. In this case, the distant stellar companion never makes a close approach to its companion's planets at any point in its orbit. After Gyrs of perturbation from the Galaxy, however, this situation can change, and the stellar companion can temporarily attain a very high eccentricity. In this state, the stellar companion begins strongly perturbing the planets of the other star as it passes very near them on each periastron passage, and this may be enough to trigger an instability within the planetary system billions of years after it formed.  Thus, very wide binaries present a unique hazard to planetary stability. Tighter binaries will have static orbits within the Milky Way's field environment. As such, they will either prohibit or allow the stability of certain planetary architectures, and this should not change over the lifetime of the system. On the other hand, most very wide binaries (assuming they have some smooth distribution of orbital eccentricities) will allow the formation of a wide range of planetary architectures. However, after the epoch of planet formation, every very wide binary is continually perturbed by the Galaxy and has the potential to evolve into a highly eccentric orbit that disrupts a once-stable planetary system.

This process of disruption has been found to be very prevalent among simulated planetary systems within very wide binaries \citep{kaib13}. An example of this type of disruption is shown in Fig.~\ref{fig:2}. Here we see that a 0.1 M$_{\odot}$ binary is placed into an initially harmless orbit about the Sun and our four giant planets. However, after $\sim$1 Gyr of evolution, perturbations from the Milky Way drive the periastron of the binary down to $\sim$100 AU, which is low enough to excite the eccentricities of the giant planets. After $\sim$3.5 Gyrs, the companion makes another low-periastron excursion, which triggers an episode of planet-planet scattering that eventually ejects Uranus. Finally, one last low periastron phase occurs after $\sim$7 Gyrs that removes Neptune from the system. Such behavior is not rare. For planetary architectures like our solar system, the effects of very wide binary companions have been studied for binary semimajor axes between 10$^{3}$ and $3\times10^4$ AU and binary masses between 0.1 and 1 M$_{\odot}$. After 10 Gyrs of evolution (the approximate main sequence lifetime of the Sun), a 0.1 M$_{\odot}$ companion has a 20-30\% probability of triggering an instability causing the loss of one or more planets. For companion masses of 1 M$_{\odot}$ this probability increases to 50--70\% \citep{kaib13}. Moreover, there is evidence suggesting that such instabilities have occurred within real planetary systems residing within very wide binaries. The most common outcome of a strong perturbation on a planetary system from a very wide binary companion is the initiation of planet-planet scattering. This process is known to eject planets and generally increase the eccentricities of surviving planets \citep[e.g., ][]{rasford96}, and Jovian mass planets within very wide binaries have been shown to have significantly higher eccentricities than Jovian mass planets around isolated stars \citep{kaib13}. This is consistent with the idea that planetary systems within very wide binaries are more prone to episodes of planet-planet scattering than planets around isolated stars. 

This process of planetary system disruption can have dramatic consequences on planetary habitability. Although very wide binary companions are most likely to directly perturb the outermost planets of a system and unlikely to directly perturb planets in the habitable zone, their perturbation to outer planetary systems can touch off global instabilities. The eccentricities excited in the outer regions of the planetary system can easily cascade into the inner regions \citep{zaktre04}. Moreover, planetary instabilities can dislodge large reservoirs of previously stable small bodies, dramatically increasing the impact rate on planets in the habitable zone \citep[e.g., ][]{gom05}. Thus, although many galactic processes have proven to have a limited influence on Earth's habitability, gravitational perturbations from the Galaxy can have dire consequences for habitable planets within very wide binaries. In addition, the studies to date of this process focus on the perturbations associated with the solar neighborhood. The effects of denser environments closer to the Galactic center can be crudely studied by scaling up the solar neighborhood's perturbations. Doing so causes even more orbital variation among very wide binaries and more planetary system disruption events \citep{kaib13}. We should expect the habitability hazards from very wide binary companions to increase toward the Galactic center and diminish further away.

\section{Galactic Habitable Zone}

Over the years many aspects of the local galactic environment's influence over the habitability of planets have been considered, and this inspired the concept of the ``galactic habitable zone'' \citep{gon01}. \citet{gon01} argued that only certain regions of the Milky Way should possess habitable planets. This work mainly focused on the distribution of metals (or atoms more massive than H or He) within the Galaxy. Namely, it was thought that Earth-mass planets would be devoid of plate tectonics (and therefore climate regulation) in metal-poor regions of the Galaxy and regions with even lower metallicity should be devoid of rocky planets completely. As we have seen, recent studies of the Earth's heating indicate that a substantial portion of the Earth's internal heat is not derived from radiogenic isotopes (whose abundances are linked to galactic metallicity). As a result, there may not be as strong of a correlation as initially thought between the prevalence of plate tectonics and the Galaxy's metallicity gradient. Furthermore, although there must be some critical stellar metallicity that prohibits the formation of rocky planets (rocky planets obviously cannot form in the complete absence of metals), recent studies suggest that low-mass planets are prevalent around stars that are substantially more metal-poor than those thought to be too metal-poor to form gas giants. Thus, the impediments that the local galactic metallicity present against the formation of habitable planets seem to be less rigid than previously supposed. 

In addition, the galactic habitable zone concept argued that the interior regions of the Milky Way, while generally metal-rich, would pose new obstacles to habitable planets in the form of comet showers, nearby supernovae, and gamma ray bursts \citep{gon01, line04}. However, as we have seen in previous sections, the most recent modeling work suggests that only supernovae within about 8 pc of Earth would erode the ozone layer by 50\%, and it is unclear if ozone depletion at this level is enough to dramatically alter Earth's habitability. Like galactic metallicity, there is likely some critical spatial density of supernova above which planets become sterilized, but this density is likely higher than initially thought. Moreover, the metallicity dependence of gamma ray bursts may confine them to the least populated regions of the modern Milky Way.

Another potential galactic hazard within crowded areas of the Galaxy are comet showers. However, our own Oort cloud has been shown to have too few objects in its inner region for comet showers to pose a major hazard to Earth's habitability. This finding is of course based on our own planetary system. The population size and structure of analogous clouds around other stars will be dependent on the architecture of the planetary system and the local galactic environment that the planetary system inhabits \citep{fern97, bras10, kaib11, lew13}. Nevertheless, there is a general trend that Oort cloud formation in denser regions (closer to the Galactic center) results in a more tightly bound cloud that is more difficult to perturb \citep{kaib11}, so the Oort clouds of stars in more crowded regions of the Galaxy may require stronger stellar passages to trigger comet showers. With so much uncertainty in the strength of comet showers, it is difficult to say where the galactic habitable zone boundary set by comet showers would lie in the Milky Way (or even if such a boundary actually exists). 

The planetary regime in which the galactic environment seems to play the most definitive role in habitability is for planetary systems within very wide binary star systems. Here denser portions of the Galaxy will more strongly perturb very wide binaries, making them more likely to evolve to a very low periastron and disrupt their planetary systems. Nevertheless, there are diminishing returns to increasing galactic density. The same perturbations that drive binaries to high eccentricity also unbind them from each other \citep{bah85, jiangtre10, kaib14}. Once binary companions are dissociated they will no longer threaten one another's planets. Thus, although very wide binary companions will be less dangerous to planetary systems (and planetary habitability) in low density regions of the Galaxy and more dangerous in high density regions, there should be no regions that completely prohibit habitable planets within very wide binary systems.

\section{Radial Migration}

As we have seen, there are many ways that a star's local galactic environment can impact the habitability of its planets. A further complication to this is the recent discovery that most stars' galactic environments may vary dramatically over their lifetimes. The seminal work of \citet{sellbin02} demonstrated that angular momentum transfer between the stars and the spiral arms of a disk galaxy can cause the stars' orbital radii to fluctuate by several kiloparsecs over Gyr time periods. Modeling the formation of a Milky Way-type galaxy, \citet{rosk08} confirmed this migration and found that, on average, stars with solar-like ages and orbits about the Galactic center can be expected to have formed 2--4 kpc closer to the Galactic center than the Sun's current position \citep{kaib11}. Moreover, recent work has shown that this radial migration is expected to flatten and add scatter to the Milky Way disk's stellar metallicity gradient, which is consistent with observations \citep[e.g.][]{loeb16, hayden15}. In this context, the most metal-rich stars at a given galactocentric distance are those that are expected to have migrated outward the most, forming from metal-rich gas nearer the Galactic center and moving outward over time \citep{loeb16}. Thus, the typical planetary system may sample a large range of galactic environments over its history, forming in more crowded regions before inhabiting more dispersed regions later. 

\section{Conclusions}

The Galaxy can influence planetary systems in a number of ways. The local metallicity of the ISM likely affects the prevalence of rocky planets and the degree of internal heating within them. However, there is evidence that stars with significantly subsolar metallicities can form rocky planets, and metallicity will have no impact on the internal heating that results from the planets' formation. After planetary systems form, they are subjected to powerful stellar encounters while the still inhabit their stellar birth clusters, but to alter the orbits of a planetary system like our own, the cluster must last 1--2 orders of magnitude longer than the typical cluster lifetime. In addition, close supernovae and GRBs can alter planetary climate through a substantial erosion of the ozone layer, but it is not clear what level of ozone depletion is necessary to cause a mass extinction or a substantial change in planetary habitability. Supernova-induced ozone depletions of 50\% or more occurring on hundred Myr timescales would require local stellar densities and star formation rates substantially higher than the solar neighborhood's. Meanwhile, the metallicity dependence of GRBs is still not understood, and this may confine them to distant, unpopulated regions of the Milky Way. Increased bombardment from comet showers can also be caused by the Galaxy, as these events are triggered by close flybys of passing field stars. However, in the case of the solar system such events are unlikely to be a cause of mass extinctions, and there is reason to believe that the structure of comet clouds would change to limit the increase in comet shower threat associated with denser portions of the Galaxy.

Perhaps the most potent hazard to habitability from the Galaxy occurs within planetary systems residing within very wide binary star systems. In these systems, the distant stellar companion dynamically links the planetary system with its larger Galactic environment. Perturbations from passing field stars and the local galactic tidal field can eventually drive the orbit of the very wide binary star through brief periods of very high eccentricity, which cause it to deliver strong perturbations to the planetary system during periastron passage, potentially destabilizing planets and altering the survivors orbital architecture. This hazard to planetary stability and habitability will generally increase with the local density of the galactic environment, but this increase will be limited by the shorter binary lifetimes associated with higher galactic densities. 

The most recent work studying the galactic environment's effects on planetary systems has altered the concept of the galactic habitable zone. Rather than there existing a well-defined thin annulus of the Milky Way's disk that can host habitable planets and vast swaths that cannot, we should view galactic habitability as a sliding scale. In general, there are regions of the Galaxy that should be somewhat more hospitable to habitable planets and other regions that are somewhat less hospitable. The exceptions to this would seem to be the very densest regions of the Galaxy, which would pose a continuous supernova hazard to planets and the most metal poor regions of the Galaxy, which may be unable to form systems with rocky planets. However, because of our lack of understanding of planet formation and the atmospheric consequences of high-energy events, we are currently unable to define the hard boundaries of the extreme areas of the Galaxy where these two effects completely inhibit habitable planets. Moreover, the radial migration of stars within spiral galaxies further complicates the situation, as stars can navigate in and out of different regions of the Galaxy. Thus, we can only conclude that the large majority of the Galaxy is likely to permit the existence of habitable worlds with varying levels of hazards. 

\begin{figure}
% For example, with the graphicx style use
\includegraphics[scale=1]{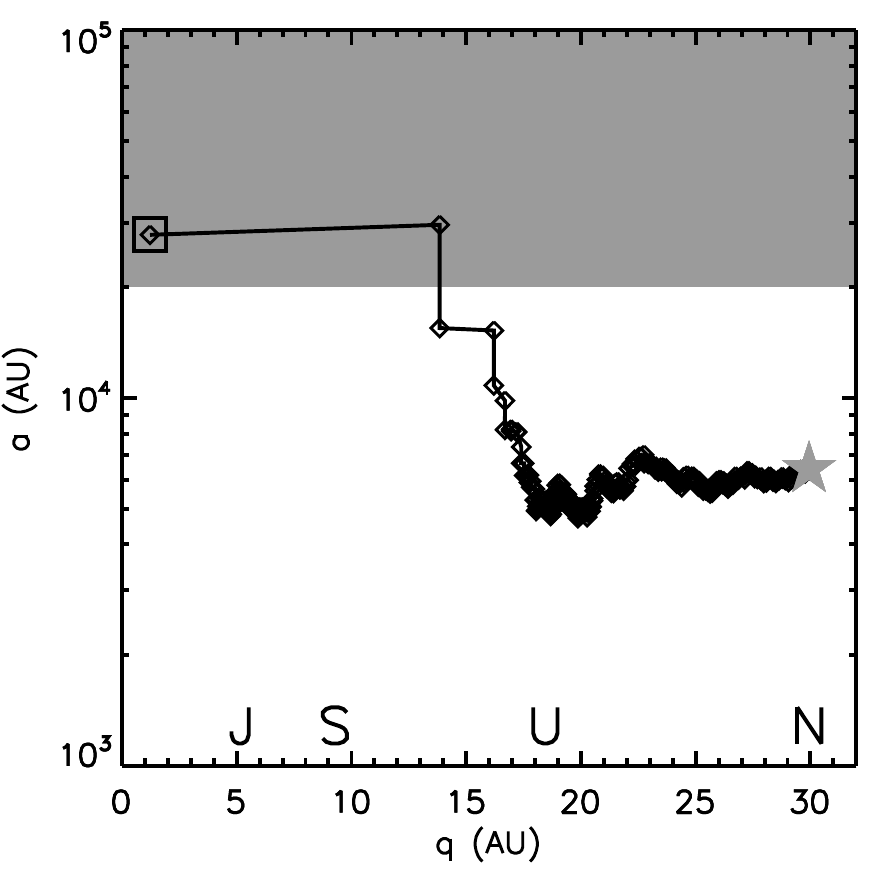}
\caption{An example of an object from the inner region of the Oort cloud evolving to a perihelion near 1 AU. The object's semimajor axis is plotted against its perihelion with a diamond data point each time it crosses the $r=35$ AU boundary of the solar system. The star marks the beginning of the orbital evolution, and the square marks the end. The approximate locations of the giant planets are marked along the perihelion axis with the planets' initials. The gray shaded region marks the region of the Oort cloud traditionally assumed to be the source of all observed long-period comets. This figure first appeared in \citet{kaibquinn09}.}
\label{fig:1}       % Give a unique label
\end{figure}

\begin{figure}
% For example, with the graphicx style use
\includegraphics[scale=1]{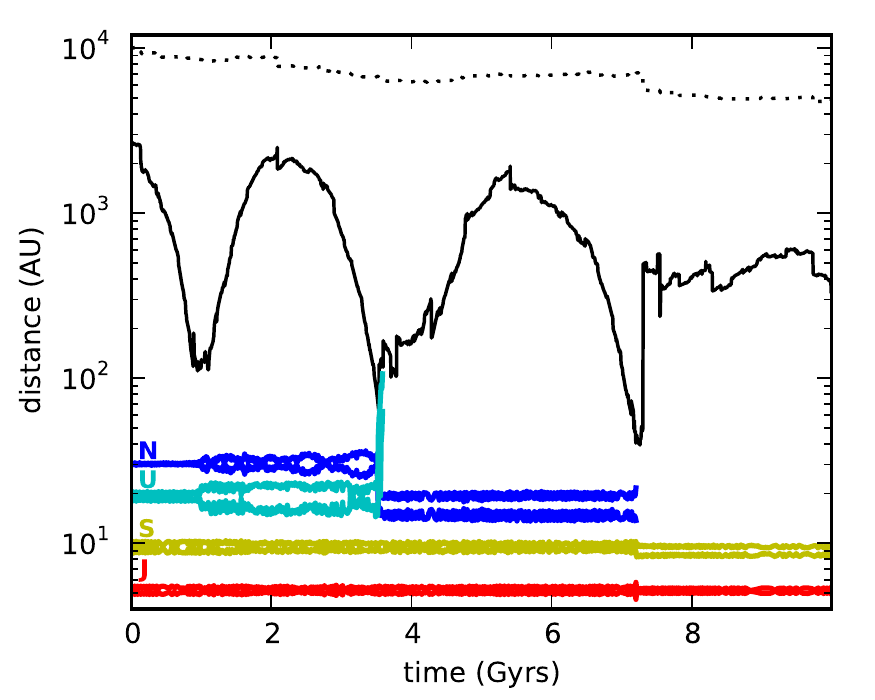}
\caption{An example of the orbital evolution of the Sun and the four giant planets when given a very wide binary stellar companion with mass 0.1 M$_{\odot}$. The stellar companion's semimajor axis and periastron are plotted against time with the black dotted and solid lines, respectively. The orbital perihelia and aphelia of Jupiter, Saturn, Uranus, and Neptune are plotted against time with red, yellow, cyan, and blue lines, respectively. This figure first appeared in \citet{kaib13}.}
\label{fig:2}       % Give a unique label
\end{figure}

\clearpage

%  IF you do NOT use bibtex, put comments before the following 2 lines
\bibliographystyle{spbasicHBexo}  %for bibtex
\bibliography{HBexoTemplatebib} %for bibtex-example

\end{document}